\documentclass[pra,showpacs,preprintnumbers,amsmath,amssymb,tightenlines,twocolumn,superscriptaddress]{revtex4}

\usepackage{graphicx}% Include figure files
\usepackage{dcolumn}% Align table columns on decimal point
\usepackage{bm}% bold math
\usepackage{epsfig}
\usepackage{stmaryrd}
\usepackage{color}

\newcommand{\bra}[1]{\langle\,{#1}\, |}
\newcommand{\ket}[1]{|\,{#1}\,\rangle}
\newcommand{\braket}[2]{\mbox{$\langle\,{#1}\, | \,{#2}\,\rangle$}}

\newcommand{\vek}[1]{\boldsymbol{#1}}

\newcommand{\rref}[1]{Ref.~\cite{#1}}

\newcommand{\eref}[1]{Eq.~(\ref{#1})}

\newcommand{\cref}[1]{chapter~\ref{#1}}

\newcommand{\Cref}[1]{Chapter~\ref{#1}}

\usepackage{ulem}  %unter- (\uline{important}) und durchstreichen (\sout{wrong})
\begin{document}

\title{Quantum or classical perception : the Imaging Theorem and the ensemble picture.} 

\author{John S.\ Briggs}
\affiliation{Institute of Physics, University of Freiburg, Freiburg, Germany} 

\email{briggs@physik.uni-freiburg.de}

\begin{abstract}
An assessment is given as to the extent to which pure unitary evolution, as distinct from environmental decohering interaction, can provide
the transition necessary for an observer to  interpret perceived quantum dynamics as classical. This has implications for the interpretation of quantum wavefunctions
as a characteristic of ensembles or of single particles and the related question of wavefunction ``collapse".
A brief historical overview is presented as well as recent emphasis on the role of the "imaging theorem" in describing quantum to classical unitary evolution.
\end{abstract}
\maketitle

\section{Introduction}
\subsection{Particle or wave or particle ensemble ?}
In the scattering of electromagnetic waves around a sharp material object, the nature of the perceived outline depends upon the resolution and sensitivity of the 
instrument. For visible light, in the case of the human eye usually a sharp outline of the object, ascribable to a ray description of the light, would be inferred. However, from measurement with an instrument able to resolve at sub-wavelength accuracy, a blurred outline corresponding to a diffraction pattern and ascribable to the wave nature of the light would be inferred. The instrument resolution is understood as the accuracy of position location. Important also is the sensitivity of the instrument, understood here as the ability or not
to register reception of a single quantum particle e.g. a photon, electron or an atom.

If the detector sensitivity is sufficient one can monitor the arrival of individual particles, in this case, photons. Their trajectories appear in a seemingly arbitrary pattern until enough photons are counted. Then the statistical distribution gradually assumes the structured form expected on the basis of the wave picture of electromagnetism. This is the wave-particle duality of light. With increasing resolution and sensitivity of the measurement, there are three levels of perception, classical ray trajectory, the wave picture and the ensemble of (quantum) particles picture. 

A similar situation arises in the wave-particle duality of matter. For an ensemble of  identical particles with a mass which is very large on an atomic scale, one assigns to their motion a unique classical trajectory so long as the resolution of, say position detection, is not itself on the atomic scale. When the mass of the particles is on the atomic scale it is necessary
to calculate the average of their motion from the wave picture of quantum mechanics. Increasing  the sensitivity to detect individual particles leads to a seemingly arbitrary pattern until the statistics are sufficient that a pattern predicted by the wave description emerges, see \rref{Bach}. Again there are three levels of perception of the ensemble; unique classical trajectory, many particles registered as a wave pattern or the statistical pattern from individual quantum particles. Which description is appropriate depends  both upon the the resolution and the sensitivity of the measurement.

Indeed the analogy between the classical wave equations of electromagnetism (Helmholtz equations and paraxial approximation) and the wave equations of quantum mechanics (time-independent and time-dependent Schr\"odinger equations) is very close mathematically. This leads to the similarity of perception
alluded to above. The semi-classical limit of quantum mechanics, used extensively below, corresponds to the eikonal approximation for electric wave propagation. This gives the quantum to classical limit 
for material particles and correspondingly the wave to beam limit of electric field propagation. The large separation between 
source and observer is used to derive the Fraunhofer diffraction formula which is in complete analogy to the "Imaging Theorem" of quantum mechanics derived in section III.

 A key element of quantum mechanics, not present for classical light, is the interpretation of the modulus squared of the wavefunction as a statistical probability. Here, two points of view have emerged. The first, to be called the ensemble picture, is that the probability describes the percentage of members of an ensemble of identical, and identically-prepared, particles having a particular value of a dynamical variable. The second, to be called the single-particle (SP) picture, considers that it is the probability with which an individual particle exhibits a given value out of the totality of possibilities. That is, on measurement the wavefunction ``collapses" into one eigenstate of the observable. The difference is that in the ensemble interpretation, only measurements on the whole ensemble are meaningful. In the SP picture meaning is assigned to a measurement on a single particle.

Here it is argued that only the ensemble interpretation of the wavefunction is tenable. However, it should be made clear from the outset that "ensemble" refers to an ensemble of 
$N$ \emph{measurements}, not necessarily $N$ particles. The initial conditions have to be identical and the wavefunction gives statistical information on the outcomes. The measurements can be simultaneous or sequential. In the case of $N$ particles the particles must be indistinguishable. This specification of ensembles of measurements is necessary since, unthinkable to the founders of quantum mechanics, experiments today can be made on trapped single electrons, atoms or molecules. Then the wavefunction gives only statistical information on a sequence of measurements in which the same particle initially is brought into the same state e.g. experiments on quantum jumps.

Furthermore, a feature that is very important but has been often neglected in the past, is the occurrence or otherwise of many-particle {\emph{good quantum numbers}} corresponding to eigenstates of some many-particle mechanical variable. This is because, for this special case, the measurement of the corresponding many-particle mechanical variable gives the same sharp value for all members of the ensemble. For the simple case of single-particle ensembles in an eigenstate, there is no difference between the SP and ensemble pictures.

The object of this work is to re-appraise, in the light of the SP and ensemble pictures, the transition from quantum to classical  mechanics  by emphasising the role of the ``Imaging Theorem" (IT) \cite{Kemble,NJP,BriggsFeagin_IT,BriggsFeagin_ITII,ITfree,Macek} in determining what an observer perceives as a consequence of the experimental resolution, sensitivity and information extraction. The IT was proved as long ago as 1937 by Kemble \cite{Kemble} whose aim was to show how a particle linear momentum vector could be measured and assigned in a collision experiment. Although largely forgotten until recently, here, following \rref{NJP}, we suggest a more fundamental consequence of this theorem.

 The IT shows that any system of particles emanating from microscopic separations describable by quantum dynamics will acquire characteristics of classical trajectories simply through unitary propagation to the macroscopic separations at which measurements are made. Specifically, the IT equates the final position wavefunction $\Psi(\vek r_f,t_f)$ at a detector at {\emph{macroscopic}} position and time, to the initial momentum wavefunction $\tilde\Psi(\vek p_i,t_i)$ at {\emph{microscopic}} position and time but, importantly, where the variables $\vek r,\vek p$ and $t$ are related by a {\emph{classical trajectory}}. This justifies the standard approach of experimentalists who use classical trajectories to trace the motion of particles from reaction zone to detector, even though the particle correlations indicate existence of a quantum wavefunction. In fact the relevance of great advances in multi-particle coincident detection \cite{Horst2} to the criteria for quantum or classical perception given here cannot be underestimated.
 
 The IT involves the connection of the {\emph{ momentum}} wavefunction in the microscopic collision zone with the {\emph {position}} wavefunction at macroscopic distance. The initial position in the collision zone cannot be defined precisely. This is completely in correspondence with recent work of Schmidt-B\"ocking et.al. \cite{Horstetal} who emphasise that momentum of particles emanating from a microscopic collision can be determined with arbitrary precision, but the initial position can never be measured with comparable precision.

 On the basis of the IT, it emerges that whether one ascribes\\ a) classical dynamics (a single trajectory analogous to a light ray),\\ b) a quantum wave description of the ensemble as a whole or \\c) single particles registered separately whose statistical distribution corresponds to a wave,\\ to the movement of material particles depends upon the precision and extent to which  the dynamical
variables of position and momentum are determined by the measurement. This is equally true for ensembles of many-particle systems involving entangled wavefunctions as it is for ensembles described by single particle wavefunctions. In the former case it is essential that the composite of several entangled particles is to be viewed as a representative member of the quantum ensemble, not the individual particles.

  The elimination of the overtly quantum effects  of entanglement and coherence as a prerequisite for the transition to classical mechanics has been ascribed to interaction with the environment  \cite{decoherence,decohere1,Schloss1,Schloss2}. It goes under the broad name of "decoherence theory" (DT). This theory is part of the wider study of open quantum systems and various amplifications of the original scheme have been proposed e.g. the ``continuous spontaneous localization" (CSL) model \cite{Ghirardi}. These approaches usually involve propagation of the quantum density matrix in time.\\
  The principal feature of such models is that the interaction with an environment leads to an elimination of the off-diagonal elements of the density matrix, which is considered a key element of the transition to classical behaviour.
  
Without doubt DT can explain many features of the quantum to classical transition but, according to DT, unitary evolution  \emph{in the system Hamiltonian alone} does not contribute to this transition.  One main aim of the present work is to show that this is not the case.
 
  Generically, according to the IT, a quantum system wavefunction or corresponding density matrix propagating in time without
 environmental interaction will develop such that the position and momentum coordinates change according to classical mechanics. In particular, the off-diagonal density matrix elements acquire oscillatory phase factors such that, except under a high-resolution measurement, they average to zero.  In this sense, the IT does not negate any predictions of DT, rather it is complementary to it. However the unitary propagation transition occurs over time and position increments which are still of atomic dimensions and thus largely obviate any additional changes to the density matrix ascribable to the environment.  
  
    An exhaustive discussion of DT with an honest appraisal of its notable successes but also its limitations is given in the reviews of Schlosshauer \cite{Schloss1,Schloss2}. It is clear that this theory is anchored firmly in the SP interpretation of the wavefunction since wavefunction collapse plays a prominent role.  \\
  
%DT can be employed also to describe certain aspects of the measurement process. In this paper discussion is confined to detection as in modern multi-particle detectors where the quantum particles, following autonomous unitary propagation through vacuum, interact, essentially instantaneously and certainly irreversibly, with a classical detector. Hence the precise details of the measurement of position are of no concern.
  
 In this work only continuum quantum states are considered of relevance in the transition to classical mechanics. Bound states and quantised internal degrees of freedom (e.g. intrinsic spin) are viewed as wholly quantum features. Furthermore, there is no discussion of the measurement process itself. In the particle detectors employed in modern experiments, the quantum particle is intercepted by a macroscopic detector involving an enormous number of atomic degrees of freedom, giving a completely irreversible transformation. The particle energy is absorbed through ionisation or photon emission in the detector  and amplified to give a recorded signal.
 
\section{Interpretation of the wavefunction}

Here a simple but sufficient interpretation of the wavefunction is applied. This involves the minimum of supposition required to explain modern multi-hit coincident detection of particles emanating from complexes of atomic dimension. The following rules are adopted in connection with the detection of moving particles.\\

1) The wavefunction always describes a statistical ensemble of  identically-prepared particles. No meaning can be ascribed to the wavefunction of a single particle.\\

2) The wavefunction  $\Psi(\vek r)$ contains information on the state of the ensemble. The wavefunction extent can be infinite or spatially confined.\\

3) The quantity $|\Psi(\vek r,t)|^2 d\vek r$ gives the probability to detect a given particle from the ensemble at position $\vek r$, at time $t$ and with a resolution $d \vek r$ (Born's rule \cite{Born}).
     The quantity $|\tilde\Psi(\vek p,t)|^2 d\vek p$. where $\tilde\Psi(\vek p)$ is the wavefunction in momentum space, gives the probability to detect a given particle from the ensemble with momentum $\vek p$ at time $t$.\\

4) When information, either partial or total, is extracted by a measurement, the corresponding part of the quantum wavefunction has been utilised and no further information can be extracted.\\

%The question of calculating probabilities from continuum scattering wavefunctions is discussed in Appendix A.

Consequent on this  ensemble view, the popular expression that a particle can also behave as a wave is redundant. What is detected is always a particle. The wavefunction simply assigns a probability amplitude that a particle from an ensemble of identical particles will be detected to have particular values of the dynamical variables. 
%This simple but sufficient interpretation of the wavefunction is inherent in the IT where unitary evolution of an ensemble according to the Schr\"odinger equation in semi-classical approximation results in a wavefunction whose co-ordinates (either space or momentum) evolve according to classical mechanics.

As will be shown in the following, the IT provides many of the features of wavefunction propagation ascribed to decoherence due to environmental interaction. However, since the propagation is unitary, classical features emerge without the need for interaction with an environment.

Wavefunction ``collapse" is a widely-accepted aspect of quantum mechanics. This concept is peculiar to the SP picture. In the ensemble picture the need to invoke collapse of the wavefunction does not arise.

\section{The imaging theorem}

The result known as the imaging theorem can be expressed in a few equations. Details of the original proof for free asymptotic motion can be found in the book of Kemble \cite{Kemble} and its generalisation for arbitrary motion, e.g. in external electromagnetic fields, in \rref{NJP}.

The propagation in time of a localised quantum state defined at time $t'$ can be written

\begin{equation}
\label{propfinal}
\ket{\Psi(t)} = U(t,t') \,\ket{\Psi(t')},
\end{equation}
where $U(t,t')$ is the time-development operator. Projecting this equation into position space gives
 \begin{equation}
\label{Psiprop2}
\Psi(\vek r,t)  = \int  K(\vek r,t; \vek r',t') \, \Psi(\vek r',t') \, d\vek r',
\end{equation}
where the function $K(\vek r,t; \vek r',t') \equiv \bra{\vek r}\,U(t,t')\,\ket{\vek r'}$ is called the space-time propagator.
The IT rests on the asymptotic large $r$, large $t$ limit when the action  becomes much greater than $\hbar$ and the propagator can be approximated by its semi-classical form \cite{Gutz}. 

\begin{equation}
 K(\vek r,t; \vek r',0) =  \frac{1}{(2\pi i\hslash)^{3/2}}\Big|{\rm{det}} \frac{\partial^2 S}{\partial \vek r\partial\vek r'}\Big|^{1/2} ~e^{i  S(\vek r,t;\vek r',0)/\hslash}
\end{equation} 
where $S(\vek r,t;\vek r',0)$ is the {\emph{classical}} action function in coordinate space and the initial time $t'$ is taken as the zero of time. 

Now it is recognised that the $\vek r'$ integral is confined to a small volume, of atomic dimensions, around $\vek r' \approx 0$, so that the action can be expanded around this point as
\begin{equation}
\label{Taylor}
S(\vek r,t;\vek r',0) \approx S(\vek r,t;0,0) +  \frac{\partial S}{\partial \vek r'}\Big|_0\cdot \vek r'.
\end{equation}
Then, using the classical relationship $ \partial S/\partial \vek r'|_0  \equiv -\vek p$, substitution in the integral \eref{Psiprop2} gives a Fourier transform and the result 
\begin{equation}
\label{IT_coor}   
\Psi(\vek r, t)  \approx (i)^{-3/2} \left(\frac{d\vek p}{d\vek r}\right)^{1/2} ~e^{i  S(\vek r,t;0,0)/\hslash} \, \tilde\Psi(\vek p,0),
\end{equation}
which is the IT of Kemble, here generalised to arbitrary classical motion.

One notes that the IT rests upon two approximations. The first is the semi-classical approximation of $K$ in \eref{Psiprop2}. However, in the integral over $\vek r'$, all possible values of $\vek r'$ contribute to the asymptotic wavefunction at $\vek r,t$. It is the recognition that the quantum wavefunction at time zero is limited to a microscopic extent, \eref{Taylor}, that associates a fixed classical momentum $\vek p$ to each final coordinate $\vek r(t)$. That is, each initial $[(\vek r' = 0), \vek p]$ value is  connected to a fixed $\vek r,t$ by a classical trajectory. For free motion the connection is simply $\vek r = \vek p\,t/m$, where $m$ is the particle mass. 

The essence of the IT result is that the position and momentum coordinates evolve classically but within the shroud of the quantum wavefunctions.

 The probability density for detection of a particle of the ensemble then is given by
 \begin{equation}
\label{VVtheorem2}
|\Psi(\vek r ( t))|^2 \approx \frac{d \vek p}{d \vek r} \, |\tilde\Psi(\vek p, 0)|^2.
\end{equation}
 Since the coordinates of the wavefunctions now conform to classical mechanics, this form has a wholly classical, statistical interpretation. An ensemble of particles with probability density $ |\tilde\Psi(\vek p, 0)|^2$, defining the
 probability of occurrence of a certain initial momentum $\vek p$, move on classical trajectories and hence the ensemble members evolve to the position probability density $|\Psi(\vek r, t)|^2$. 
 
 The factor $d \vek p/d \vek r$ is the \emph{classical} trajectory density of finding the system in the volume element $d\vek r$ given that it started with a momentum $\vek p$ in the volume element $d \vek p$ (see Gutzwiller \cite{Gutz}, chap.\ 1). Quantum mechanics provides the initial  ensemble momentum distribution located at a microscopic distance
 $\vek r' \approx 0$. Each element of the initial momentum wavefunction is then imaged onto the spatial wavefunction  at large distance $\vek r$, where the coordinates are related by classical mechanics.
 
 That is, from \eref{VVtheorem2} one has the asymptotic equality of probabilities in initial momentum space and final position space, i.e.
  \begin{equation}
\label{VVtheorem3}
|\Psi(\vek r(t))|^2~ d \vek r  =  |\tilde\Psi(\vek p, 0)|^2~d \vek p
\end{equation}
 This shows that the loci of points of equal probability of particle detection are classical trajectories. Nevertheless, according to \eref{IT_coor},  the wavefunction remains intact. 
 
 Clearly, the IT can only be interpreted in the ensemble picture. The wavefunction spreading corresponds to the natural divergence of  an ensemble of classical trajectories of differing initial momentum emanating from a microscopic volume and being detected after traversing a macroscopic distance. Nevertheless, estimates of the $\vek r$ and $t$ values at which the semi-classical approximation becomes valid (\rref{NJP}) show that this occurs for values which are still microscopic, typically only tens of atomic units, the precise value dependent upon particle masses and energies.
 
It is to be emphasised that the IT describes classical evolution of the wavefunction variables and the transition to this property arises from unitary quantum propagation i.e. the transition to classical behaviour is autonomous; external interactions are unnecessary. This justifies a routine assumption of experimentalists that one can use classical mechanics to trace a trajectory back from a point on the detector to the quantum reaction zone and is valid even for light particles such as electrons.

The consideration of the quantum to classical transition from a more mathematical viewpoint, so-called semi-classical quantum mechanics, began with the early WKB
approximations and Van Vleck's work on time propagators \cite{VV}. It was formulated initially for scattering theory, for example by  Mott and Massey \cite{MottMass}, by Ford and Wheeler \cite{Wheeler} and by Brink \cite{Brink}.
A completely general theory emerged later in the work of  Berry and Mount \cite{Berry} and of Miller \cite{Miller}, for example. Major contributions made by Gutzwiller are to be found in \rref{Gutz}.

In  semi-classical scattering theory one examines the transition to a classical cross-section which occurs when the collision energy is much greater than the interaction energies of the collision complex, see for example, \cite{RostHeller,Rost}. This is to be contrasted with the IT in which quantum systems of atomic dimension are described fully by quantum mechanics but  the transition to macroscopic distances by the semi-classical approximation. Then the semi-classical description is valid for all energies, after distances are traversed such that the classical action far exceeds $\hbar$. This is the autonomous aspect of the quantum to classical transition.

\section{The quantum to classical transition}

\subsection{Historical context}

The question of the transition from quantum to classical mechanics in the motion of particles is as old as wave mechanics itself. In the SP picture it is required that in the classical limit the wavefunction of a single particle describes a classical trajectory i.e. a narrow wavepacket. In the ensemble ppicture, the limit is, as described by the IT, that the wavefunction describes an ensemble of particles following classical trajectories.

 Schr\"odinger, immediately following his invention of wave mechanics in a sequence of papers in 1926, investigated the classical limit of wave mechanics. In a paper \cite{Schr} entitled "On the continuous transition from micro- to macro-mechanics" he gave an example of how a packet of waves describing the harmonic oscillator can move in such a way that the displacement of the wavepacket as a whole follows the well-known classical
dynamics of the one-dimensional harmonic oscillator. In this calculation Schr\"odinger repeatedly draws the analogy of superpositions of oscillator eigenfunctions to wavepackets formed from classical normal modes on an oscillating string. 

The important point to note here is that Schr\"odinger was seeking, through the wave equation, to represent
a {\it{single particle}} as a packet of quantum waves which is so localised in space that it can be perceived as a classical particle. Nevertheless he recognized the limitations of his model, pointing out, for example, that a non-dispersive packet can only be built from \emph{bound} eigenfunctions and any admixture of continuum states will result in an expanding wavepacket as in the optical case.

 This latter point was taken up by Heisenberg \cite{Heisenberg1} in a lengthy paper on the interpretation of the new quantum mechanics and its relation to classical mechanics. In a section also called ``the transition from micro- to macro-mechanics", Heisenberg criticises the relevance of bound states in connection with classical mechanics. To illustrate the difficulty with continuum motion Heisenberg showed that an initial Gaussian wavepacket moving freely will spread in space as a function of time and so cannot represent a single material particle.

A more precise demonstration of the classical aspects of quantum motion can be traced back to 1927 in a paper by Kennard \cite{Kennard}.  Kennard showed that the {\emph{centroid}} of quantum ``probability packets" moves according to classical mechanics. In retrospect, Kennard probably deserves recognition for the ``Ehrenfest" theorem, but perhaps this is denied him since he couched his proof in the language of matrix mechanics, whereas Ehrenfest  \cite{Ehrenfest} used Schr\"odinger wave mechanics.
 
Kennard's paper, although little quoted, is a very important landmark in the development of the meaning of the wavefunction.
Interestingly, this is one of the last papers to utilise predominantly the Born, Heisenberg, Jordan \cite{BHJ} theory of matrix mechanics. Kennard  defines a ``probability amplitude" $M(q)$ for a variable $q$ in matrix mechanics, which is later shown to be equivalent to the Schr\"odinger wavefunction $\psi(q)$.  

He considers the motion of ``probability packets" and shows that, for the cases of free motion or motion in constant electric or magnetic fields, the centroid obeys classical mechanics. 

As perhaps the first to emphasise the ensemble picture, Kennard shows that Heisenberg's ``proof" of the uncertainty principle is properly formulated as the statistical spread of momentum and position measured on an ensemble of identical systems. The spread, for the particular case of a free wavepacket, is calculated using the probability $MM^*~dq$ which is identical to Born's probability interpretation of the Schr\"odinger wavefunction.
 
 As mentioned above, the case of free motion had been solved already by Heisenberg
 \cite{Heisenberg1} who showed that a Schr\"odinger free wavepacket spreads in time. Kennard, although he shows that his probability amplitude $M$ is the same as a Schr\"odinger wavefunction $\psi$, uses this spreading as an argument against the superiority of Schr\"odinger wave mechanics with respect to matrix mechanics. 
 
  Kennard raises objections to the Schr\"odinger wave equation by pointing out that a spreading wavefunction of an electron must correspond to a spreading of charge density. Note that here, in contrast to his view of the $M$ of matrix mechanics, in interpreting Schr\"odinger's $\psi$, Kennard is assuming that the SP picture applies to this wavefunction. Then he points out that a detection of the electron must localise its full charge at a point. Hence, because of the measurement, the original diffuse wavepacket ``loses any further physical meaning" and must be replaced by  ``a new, smaller wavepacket". Kennard is using the necessity, in the particle picture, to invoke a ``collapse of the wavefunction"  as an argument against the use of a Schr\"odinger wavefunction.
 
 Following this objection to the collapse scenario, Kennard then advances the ensemble interpretation of the probability amplitude of matrix mechanics. He writes ``the wavepacket spreads, for example, like a charge of shot, in which each pellet describes a trajectory dependent upon its initial position and motion and the whole charge spreads in time as a consequence of differences in these initial conditions", precisely as described by the IT \eref{VVtheorem2}. In the ensemble picture, as distinct from the SP picture, there is no problem with the spreading of the wavepacket. Classical particles with different initial momenta will spread out as they move from micro- to macroscopic distances.

 Ehrenfest's paper  was published a few months after Kennard's.  Apparently, the clarification of the connection of quantum to classical mechanics received an enormous boost with this publication. Ehrenfest used the Schr\"odinger equation to prove
the theorem showing that quantum position and momentum {\it{expectation values}} obey a law similar to Newton's law of classical mechanics. 
In one dimension it is expressed as
\begin{equation}
m\frac{d^2\langle x\rangle}{dt^2} = \int dx~\Psi\Psi^* \left(-\frac{\partial V}{\partial x}\right) =  - \langle\frac{\partial V}{\partial x}\rangle
\end{equation}
As often remarked, however, this is not Newton's Law which would require $- \partial \langle V\rangle/\partial x$ to appear on the r.h.s..
 However, it turns out that for the cases $ V = a, V=ax$ and $V = ax^2$, where $a$ is a real constant, the theorem is the same as Newton's law. The spreading of wavepackets remains a problem however. If the wavepacket occupies a macroscopic volume of space, little meaning can be attributed to an average position.
 Also, for all other potentials with terms higher than quadratic one does not have motion according to Newton's law. Hence, for  these two reasons and despite its appealing form, in general Ehrenfest's theorem cannot be considered as describing the transition to classical mechanics, as emphasised by Ballentine \cite{Ball1,Ball2}.

Mindful of Heisenberg's proof of free wavepacket spreading, Ehrenfest is careful to stress that, within the particle picture, the motion of the mean value according to Newtonian mechanics is 
meaningful only ``for a small wavepacket which remains small (mass of the order of 1gm.)". Clearly he was thinking of a single particle described by a small wavepacket.
The ideas that narrow wavepackets and Ehrenfest's theorem embody the nature of the quantum to classical transition for a single particle, pervade most elementary text books on quantum mechanics even today.

 \subsubsection{After 1927}
 
 It is interesting, although understandable in the first years of quantum and wave mechanics, that the SP and ensemble pictures are continually confused. This applies not only to Kennard, as outlined above,
 but also to Heisenberg and Schr\"odinger themselves. In discussing the uncertainty principle, Heisenberg describes exclusively measurements on a single particle, as is discussed in great detail by Schmidt-B\"ocking et.al. \cite{Horst2}. This is despite the Kennard paper quoted above and above all Robertson's proof \cite{Robertson} of the uncertainty principle. Both papers make clear that the spread of measured values of a variable refers to an ensemble statistical spread and not the uncertainty in measuring that property on a single particle. Similarly, Schr\"odinger,  although a confirmed advocate of the SP picture, still admits the validity of Born's  statistical interpretation and the necessity to consider a sequence of measurements, see the discussion of Mott's problem given below.

 Although the Ehrenfest Theorem and narrow wavepackets  are used as the classical limit in many elementary text books, reminders have been given continually since 1927 of the problems involved with this picture and the essential interpretation of a wavefunction as representing an ensemble and not a single particle.
 
 Kemble in 1935 \cite{Kemble1}, comments that the interpretation of quantum mechanics ``asserts that the wavefunctions of Schr\"odinger theory have meaning primarily as descriptions of the behaviour of (infinite) assemblages of identical systems similarly prepared". 
 
 Writing in 1970, Ballentine \cite{Ball1} advances several arguments ``in favour of considering the quantum state description to apply only to an ensemble of similarly prepared systems, rather than supposing as is often done, that it exhaustively represents a single physical system".
  In a scholarly essay in 1980, on the ``Probability interpretation of quantum mechanics", Newton \cite{Newton} emphasizes that `` the very meaning of probability implies the ensemble interpretation".
  
 In 1994, Ballentine et.al. \cite{Ball2} examined the Ehrenfest theorem from the point of view of the quantum/classical transition and concluded that
``the conditions for the applicability of Ehrenfest's theorem are neither necessary nor sufficient to define the classical regime."
Furthermore, in connection with the ensemble or SP pictures they concluded that ``the classical limit of a quantum state is an ensemble of classical orbits, not a single
classical orbit."

\section{Consequences of the IT and the ensemble picture.}

In this section three classic problems of quantum theory are analysed briefly within the IT and related ensemble picture. The problems are the subject of countless papers and the ensemble aspects have ben discussed before. However, the consequences of the IT illuminate further the simplicity of the ensemble explanation. Then the reconciliation of the classical trajectory aspect of IT with the quantum interference effect is presented.

\subsection{The Schr\"odinger Cat}

The mere posing of this question by Schr\"odinger \cite{Schr1} attests to his adherence to the SP interpretation of the wavefunction. As has been observed earlier, in the ensemble picture
the interpretation is trivial. Since the wavefunction applies to many observations, one finds that half the cats are alive and half are dead. No meaning can be attached to the observation of a single cat,
unless successive measurements are made over time and feline re-incarnation is allowed.

In the same paper,  Schr\"odinger comments on the apparent problem that radioactive decay described by a spherically-symmetric wave does not lead to uniform illumination
of a spherical screen but rather to individual points which slowly are seen to be uniformly distributed. However, although he states that "it is impossible to carry out the experiment with a single radioactive atom" he does not concede that this requires an ensemble interpretation of the wavefunction. This is precisely the problem of Mott which is considered next.

\subsection{The "Mott Problem" of track structure}
 
One of the oldest "problems" of the interpretation of a wavefunction for material particles is that posed by Schr\"odinger \cite{Schr1} and addressed in 1929 by Mott \cite{Mott}. This is one of the 
most striking examples of erroneously assigning a wavefunction to a single particle. Mott remarked,

``In the theory of radioactive
disintegration, as presented by Gamow, the $\alpha$-particle is represented by a
spherical wave which slowly leaks out of the nucleus. On the other hand, the
$\alpha$-particle, once emerged, has particle-like properties, the most striking being
the ray tracks that it forms in a Wilson cloud chamber. It is a little difficult
to picture how it is that an outgoing spherical wave can produce a straight
track ; we think intuitively that it should ionise atoms at random throughout
space."

Mott presents a detailed argument based on scattering theory to argue that only atoms lying on the same straight line will be
ionised successively by an $\alpha$-particle emitted in a spherical wave. Although Mott repeatedly refers to the {\emph{probability}} 
of ionisation he interprets the wavefunction as applying to a single  $\alpha$-particle. 

However, according to the IT and the ensemble interpretation
the proof of Mott is completely superfluous. There is absolutely no mystery attached to ``how it is that an outgoing spherical wave can produce a straight
track".  This apparent dichotomy of wave mechanics is explained by the dual nature of the semi-classical wavefunction of \eref{IT_coor}; quantum wavefunction with classically-connected coordinates. Each coordinate of the initial momentum wavefunction corresponds to a specific momentum and therefore to a specific position $\vek r(t)$ along the classical trajectory. The spherical $S$ wavefunction applies to the ensemble as a whole and specifies equal probability of emission in all directions, i.e. uniform distribution of $\vek p$ on the unit sphere. Each $\alpha$-particle
is launched with a momentum $\vek p$ and this coordinate of the initial momentum wavefunction, according to the IT, follows a {\emph{straight line}} classical trajectory. Hence
it is obvious that only atoms lying along this trajectory can be ionised and the usual straight track in the cloud chamber is observed. 

 This is a prime example of the principle that what one perceives, in this case directed motion, a classical trajectory, or spherically uniform distribution, a quantum probability, depends upon the nature and precision of the experiment.
 
\subsection{Entanglement and Wavefunction collapse}

   That wavefunction superposition applies to an ensemble is made clear also by the process of radioactive decay discussed above. Although usually thought of in the time domain, the stationary picture is simpler. An ensemble of nuclei is described by a superposition of the state of a bound nucleus and a state of two  separated product nuclei
 at the same total energy. The intrusion of a measuring device simply detects which state a given member of the ensemble occupies. The absence of a signal in a measuring device denotes undecayed state and a signal denotes a decay. The half-life is interpreted from a sequence of measurements on the ensemble. It is not a property of a single nucleus, although colloquially the half-life is often so ascribed. This aspect is emphasised particularly in the very clear exposition of Rau \cite{Ravi}.

 The paper of Einstein, Podolsky and Rosen \cite{EPR}, whose result often is referred to as the ``EPR paradox", has been the subject of an enormous number of works on the subject of reality, action at a distance etc.. Throughout the EPR paper appears the SP viewpoint of a partial wavefunction describing an independent particle. 
 
 Already in the first replies to the EPR paper, by Schr\"odinger \cite{Schr2} and Bohr \cite{Bohr}, it was pointed out that it is essential to consider the {\emph{two-particle}} commuting operators,
 ignored by EPR. Nevertheless the reply papers did not apply these considerations directly to the EPR entangled wavefunctions.

Here we infer the ensemble picture and show that the recognition of good two-particle  quantum numbers is essential. Then, in the pure states considered in EPR, a  good quantum number ensures that every pair of the ensemble will give the same value of the corresponding two-particle property upon measurement.

EPR consider a two-particle eigenstate written in the entangled form
\begin{equation}
\label{entangp}
\Psi(x_1,x_2) = \int \psi_p(x_2) u_p(x_1)~dp
\end{equation}
where 
\begin{equation}
\label{pfns}
u_p(x_1) = e^{\frac{i}{\hbar}px_1}~~~  {\rm{and}}~~~\psi_p(x_2) = e^{-\frac{i}{\hbar}p(x_2 - x_0)}
\end{equation}
are eigenfunctions of one-particle operators $p_1,p_2$ with eigenvalues $p$ and $-p$ respectively. The constant $x_0$ is arbitrary.
Note that the single-particle momentum $p$ can take any value. 

The $p$ integral in this equation can be carried out to give
\begin{equation}
\label{entangx}
\begin{split}
\Psi(x_1,x_2)& = 2\pi\delta(x_1 - x_2 + x_0) \\&= 2\pi \int \delta(x_1 -x)\delta(x - x_2 + x_0)~dx
\end{split}
\end{equation}
which is an entangled state in position space. However, again, all $x$ values are possible.

Thus it has been shown that one and the same two-particle function can be expanded in terms of eigenfunctions of observables of particle $2$, in this case $p$ and $x$, which do not commute. 

As shown by Schr\"odinger \cite{Schr2} and Bohr \cite{Bohr}, the conserved quantities emerge from a transformation to relative and centre-of-mass (CM) coordinates for equal mass $m$ particles. We define relative $x_r$ and CM position $X$ as
\begin{equation}
x_r = x_1 - x_2~~~~~{\rm{and}}~~~~X= (x_1 + x_2)/2
\end{equation}
and correspondingly relative and CM momenta
\begin{equation}
p_r = (p_1 - p_2)/2~~~~~{\rm{and}}~~~~P_{CM} = p_1 + p_2
\end{equation}

Immediately one sees from \eref{entangx}, that $\Psi(x_1,x_2)$ {\emph{is}} an eigenfunction of the relative position coordinate $x_r = x_1 - x_2$ with eigenvalue $x_r = -x_0$. Similarly, from \eref{entangp} with \eref{pfns}
one sees it is simultaneously an eigenfunction of CM momentum $P_{CM}$ with eigenvalue zero. This is in order since these two operators commute. However it is readily checked, as must be, that $\Psi(x_1,x_2)$
{\emph{is not}} an eigenfunction of $X$ or $p_r$ since these do not commute with $P_{CM}$ and $x_r$ respectively. 

In summary, the two-particle wavefunction of EPR fixes
the CM momentum at zero and the relative position of the particle pair is equal to $-x_0$. This is the only information in the two-particle wavefunction.  One has, however, the clear requirement that the two-particle wavefunction should propagate intact to the detectors. In any measurement the two corresponding two-particle observables have the same precise value for all members of the ensemble of pairs.

Now one has two possible scenarios characterising entanglement.

a) If one knows the two-particle good quantum numbers in advance e.g. by selection rules on state preparation, then
the determination of the single-particle momentum to be $p = p_1$  fixes $p_2 = -p_1$. Similarly measurement of $x_1$ fixes $x_2 = x_1 + x_0$.

b) If one does not know the quantum numbers in advance, one must perform measurements on many two-particle systems {\emph{in coincidence}}. Then one can ascertain by experiment that, for all ensemble members, whatever the measured values of $p_1$ and $x_1$, one measures always $p_2 = -p_1$ and $x_2 = x_1 + x_0$.

 The measured two-particle eigenvalues are sharp but the single-particle $p$ values have a distribution of probability predicted by projection of the one-particle probability amplitude out of the two-particle wavefunction. 

Both scenarios require non-local information. The measurement in b) requires communication between the two separated detectors to ensure coincidence. In case a) only one detector is required but the non-local information is in the knowledge of the two-particle quantum numbers which are conserved for all particle separations.

Note that the specification of {\emph{two-particle} conserved observables allows one to assign precise values to both non-commuting {\emph{one-particle} observables.

The simultaneous fixing of position and momentum becomes apparent within the IT, if as is normal, detection is made at large distances from the volume from which the correlated pair are created. according to the IT there is a classical connection between position and initial momentum for detection of particles $1$ and $2$ at times $t_1$ and $t_2$ respectively. Then the space wavefunction can be written
\begin{equation}
\label{entangIT}
\Psi(x_1,x_2) \propto \tilde\Psi(p_1, p_2 = -p_1).
\end{equation}
 In particular the IT gives the classical relation
\begin{equation}
x_1 = p_1t_1/m~~~~{\rm{and}}~~~~x_2 = -p_2t_2/m
\end{equation}
so that from the second conservation law $x_2 = x_1 + x_0$ one has the restriction
\begin{equation}
x_0 = -(p_1t_1 +p_2 t_2)/m.
\end{equation}
 Single-particle $x$ and $p$ can be measured simultaneously with sub-$\hbar$ accuracy, see  \rref {Horst}.

A striking manifestation of such entanglement, which has been well-studied in experiments, is the full fragmentation of the helium atom by a single photon.
This example is given since it comprises both the momentum entanglement of EPR {\emph{and spin entanglement}} in a pure two-electron state. Furthermore, from the IT, the electrons can be assigned classical trajectories \emph{within the two-electron quantum wavefunction}. This is not a ``Gedankenexperiment" but a real measured system \cite{BrSchm}.
 
The two electrons emerging can be detected in coincidence and occupy a $^1P^o$ two-electron continuum state (this means their state is a spin singlet, has total orbital angular momentum one unit and odd parity). A selection rule \cite{MaulBr} says that electrons of the same energy cannot be ejected back-to-back i.e at $180^\circ$ such that $\vek p_1 = -\vek p_2$. That is, the \emph{two-electron} state has a node for the EPR configuration as the coincidence experiments confirm.

 If one  of the electrons is left undetected a counter will register electrons of a given energy at a particular angle. However, if
 a detector diametrically opposed is switched on to detect electrons of the same energy in coincidence, the counts in both detectors will be zero. This coherent state can be made incoherent by switching off one of the detectors when electrons will be measured again. The essence is that this pure effect of wavefunction entanglement is evident, even though according to the IT, the electrons are moving on classical trajectories after they exit the reaction zone with well-defined momenta.\\
 
In interpreting the wavefunction, as in EPR, it is crucial that the ensemble is viewed as an ensemble of {\it{two-electron}} systems. This two-electron wavefunction is the single quantum entity and it must be transmitted to the macroscopic detection zone unchanged. Then there is no wavefunction interpretation problem with the ensemble picture. The wavefunction node says that the total ensemble simply has zero probability that a given member (pair of electrons) will be emitted in the forbidden configuration. \\
The coincidence detection of both position and momentum extracts the information from the wavefunction of  the ensemble  of two-electron states. The non-coincident detection of electrons extracts information on the ensemble of single electrons. The effect of entanglement is non-local simply because the two-electron wavefunction is non-local.\\

 \subsection{Quantum interference}
 
 The explanation of interference patterns in terms of semi-classical wavefunctions and the underlying classical trajectories has been given in great detail by Kleber and co-workers
\cite{Manfred} and will not be repeated here. Based upon the IT (see eq.(1) of Kleber \cite{Kleber1994}), their theory is used to interpret experiments such as those of Blondel et.al.\cite{Blondel}. Here the ``photoionisation microscope" exhibits interference rings of electrons ionised from a negative ion in the presence of an extracting electric field. In the semi-classical explanation electrons can occupy two classical trajectories. Either they proceed directly to the detector or, initially they are ejected moving away from the detector but are turned around in the electric field. The imaging of the spatial wavefunction squared is obtained by detection on a fixed flat screen i.e. the position only of electrons is detected. Then an interference pattern from the two trajectories is observed. \\
However, were the vector position {\it{and}} vector momentum of the electrons to be observed, that would correspond to a ``which way " determination and the perception would be of
two distinct classical trajectories. Interestingly, as distinct from entanglement, in this case it is a lack of information which gives rise to wave perception. Blondel et.al. \cite{Blondel}
remark also that for ionisation from neutral atoms the interference rings are there but are too small to be detected, again showing that perception depends upon resolution.

 \section{The Imaging Theorem and Decoherence Theory: IT and DT.}
 
 As stated in the Introduction, the suppression of state superposition, entanglement and interference through environmental interaction can be seen as a requirement on the way to a classical limit of quantum mechanics and has come to be known as ``decoherence theory" (DT).  It is viewed as a universal phenomenon, extending even to the classical limit of quantum gravity \cite{Halli,Kiefer} (for an interesting discussion see \rref{Rugh}).
 In the following the transition from quantum to classical perception is discussed.
  
 There is an enormous literature on DT and alternative models such as lead to ``spontaneous localisation" due to stochastic interaction. Space does not permit a discussion of the many and varied aspects of 
 these theories, so here consideration is given to those features relevant to the quantum to classical transition embodied in the IT and to the SP or ensemble interpretation of the wavefunction. 
 
 The essence of DT is given in the famous paper of Zurek \cite{decohere1} and in more detail in the reviews of Schlosshauer \cite{Schloss1,Schloss2}. A more exhaustive treatment with discussion of the $\hbar$ dependence of the environmental interaction terms is to be found in the stochastic Schr\"odinger equation approach \cite{Walter}. Here the simpler original density matrix version of Zurek \cite{decohere1} is sufficient as illustration.
 
 The basic mechanism of DT by which certain quantum aspects are eliminated is quite straightforward, accounting for the universality of this phenomenon. In the simplest case presented in Ref.~\cite{Schloss1}, a one-dimensional two-state quantum system $S$, with wavefunctions $\psi_n$, is assumed to become entangled with an ``environment"  with corresponding wavefunctions $E_n$. Limiting to two-state quantum systems, the ensuing entangled state vector is 
 \begin{equation}
\ket{\Psi} = \alpha \ket{\psi_1} \ket{E_1} + \beta \ket{\psi_2} \ket{E_2}
\end{equation}
and gives a total density matrix $ \rho = \ket{\Psi}\bra{\Psi}$. According to Ref.~\cite{Schloss1}, ``the statistics of all possible local measurements on $S$ are exhaustively encoded in the reduced density matrix $\rho_S$", given by
\begin{equation}
\label{denmatrix}
\begin{split}
\rho_S~ = ~&  Tr_E \rho~ =  |\alpha|^2 \ket{\psi_1}\bra{\psi_1} + |\beta|^2 \ket{\psi_2}\bra{\psi_2} \\
& + \alpha\beta^* \ket{\psi_1}\bra{\psi_2} \braket{E_2}{E_1} + 
\alpha^*\beta \ket{\psi_2}\bra{\psi_1} \braket{E_1}{E_2}.
\end{split}
\end{equation}
Then a measurement of the particle's position is given by the diagonal element,
 \begin{equation}
 \label{denmatrix2}
 \begin{split}
\rho_S(x,x) =  & |\alpha|^2~|\psi_1(x)|^2 + |\beta|^2~|\psi_2(x)|^2\\
& + 2 {\rm{Re}} [\alpha\beta^*\psi_1(x)\psi_2^*(x)\braket{E_2}{E_1}]
\end{split}
\end{equation}
where ``the last term represents the interference contribution". The assumption of DT is that in general the states of the environment are orthogonal and so the interference term disappears. More importantly, from \eref{denmatrix} the off-diagonal terms disappear and one has a diagonal density matrix only. From \eref{denmatrix2} this has two ``classical" terms interpreted as classical probabilities.\\

 A slightly different model is adopted in \rref{decohere1} in that the two states  comprising the system $S$ are taken as two spatially-separated Gaussian wavefunctions. The corresponding  system density matrix exhibits four peaks. This density matrix is propagated in time subject to a temperature-dependent environment interaction. The result is to give a  density matrix of diagonal form with only two  peaks along the diagonal. 
 
 In this case the decoherence reduces the off-diagonal elements to zero and the diagonal term does not contain the ``interference" contribution since the Gaussians do not overlap. This  removal of coherence between different spatial parts of the wavefunction is considered to correspond to the emergence of classicality. 
  In connection with the classical transition Schlosshauer writes \cite{Schloss2} ``the interaction between a macroscopic system and its environment will typically lead to a rapid approximate diagonalisation of the reduced density matrix in position space and thus to spatially localised wavepackets that follow (approximately) Hamiltonian trajectories". This following of classical trajectories however, is not proven in detail. 
 
 Implicit is the SP picture in which the diagonal elements represent narrow wavepackets giving classical behaviour via Ehrenfest's theorem. The ultimate spreading of these wavepackets is not considered, although suitable environmental interaction can lead to the wavepackets remaining narrow. In short, the transition to classicality is viewed as an elimination of quantum coherence effects and the vital feature of the emergence of classical dynamics according to Newton not shown.
 
In appendix A, following the example of \rref{decohere1}, the free \emph{unitary} propagation of two, initially narrow, Gaussian wavepackets within the IT is calculated. It is shown that, under low detector resolution, the density matrix also assumes the diagonal form
\begin{equation}
\rho(x,x,t) =  \frac{1}{\sqrt\pi \eta(t)} ~(e^{-(x - X_1)^2/\eta^2} + e^{-(x - X_2)^2/\eta^2})
\end{equation}
where $X_1,X_2$ are the centres of the wavepackets and the time-dependent width is $\eta = \tilde\sigma t/\mu$, for initial width $\tilde\sigma$ and particle mass $\mu$. Hence, the intrinsic spreading of the wavepacket with time emerges as expected in the ensemble picture. In this picture there is no problem of interpretation of the two probabilities; $50\%$ of the ensemble members will be detected near to $X_1$ and $50\%$ near to $X_2$. Wavefunction collapse is unnecessary. Most important however, in the IT, the  propagation of the co-ordinates of the diagonal density matrix is according to {\it{classical}} mechanics. Nevertheless, if the resolution is on the microscopic scale then interference and manifestations of quantum propagation resulting from finite off-diagonal elements can be detected. Just as in optics, the perception of particle trajectory (ray) or wave is decided by the sharpness of vision.

The study of collision complexes in nuclear, atomic and molecular physics has long been concerned with the questions of measurement of interference and entanglement effects \cite{Crowe,FanoJoe,Blum}. 
Coincidence detection of several collision fragments in entangled states are performed with increasing sophistication (see, for example, \rref{Horst2,Helm2014}). In line with the IT, classical motion of the collision fragments  outside the reaction zone is shown to be appropriate. Nevertheless quantum coherence is preserved showing that environmental decoherence does not occur in such experiments.

 The degree of decoherence assigned to a many-body entangled state depends upon which particles are not observed or even which dynamical properties are observed and which are not. Coherence can be fully or partially removed according to the experiment. In the language of the experimentalist, either one registers the ``coincidence" spectrum or the ``singles" spectrum. Again this illustrates that perception of quantum effects depends upon the measurement. Non-detection of collision variables corresponds to a partial trace of the full density matrix, as in DT.

\section{Conclusions}

The imaging theorem corresponds only to the ensemble interpretation. According to the IT, an initial momentum distribution decides the spatial wavefunction at macroscopic distance. This corresponds to an ensemble of  classical particles with the same initial momentum distribution.
Each particle appears to move along a classical trajectory to be registered at well-defined position at a distant screen. The loci of points of equal probability are the classical trajectories but the probability is given by the quantum position wavefunction. 

Indeed, all collision experiments support the ensemble picture. One counts many particles at different locations and times on a detector and so builds an image of the initial momentum distribution. Particularly striking in this respect is the observation of the gradual assembly of an interference pattern. Using electron diffraction through a pair of slits, it has been shown \cite{Bach} that the wave interference pattern is built up slowly by registering many hundreds of hits of individual electrons on a detector screen. That it is the ensemble of hits at the detector that builds up the wave interference pattern and not a single particle carrying the wave, has been demonstrated convincingly also in \rref{Vienna}.

The SP picture is that the wavefunction, extending over macroscopic distance, represents the potential detection position of each single electron. The detector is required to instigate decoherence leading to instantaneous wavefunction collapse (from macroscopic to microscopic extent) and the electron being registered at a single localised point on the detector. Again, one is faced with the dilemma of Kennard in understanding such a transition.

It has been shown that;\\

1) The IT preserves the quantum wavefunction but the momentum and position \emph{coordinates} change in time according to classical mechanics.\\

 2) as a result of the IT, unitary evolution of quantum systems, even over microscopic distances, leads to perception of an ensemble of particles as following classical trajectories. \\
 
 3) Standard measurement techniques, either on single or multiple particles, can lead to perception or otherwise of the quantum properties
 of interference and entanglement according to the information registered. The inference of classical or quantum behaviour depends ultimately upon the resolution and detail of the measurement performed.\\

Without environment influence, within the IT, unitary evolution of quantum systems results in effective decohering effects.
This ``decoherence" is of a different nature than in DT. It occurs due to cancellation of oscillating terms of different phase, which leads to non-resolution of oscillatory terms in the propagation of the density matrix to macroscopic times, as in \eref{rhodiag} and \eref{rhooffdiag}.
 Hence, lack of sufficient resolution results in effective decoherence although paradoxically it arises from the very terms, oscillatory phase factors, which are the hallmark of quantum coherence in the wavefunction.
 
The preservation of the wavefunction can lead to interference. However, the perception of interference patterns, or not, again depends upon the nature of the measurement performed. The observation of interference patterns implies that, although resolution is high, incomplete information as to the different trajectories encoded in the wavefunction variables is extracted by the  measurement. That is, a ``which way" detection is not performed. Then, whether one perceives quantum or classical dynamics depends simply upon the precision of the measurement performed and the amount of information extracted from the wavefunction. This is all in close analogy, both physically and mathematically, to the optical case of perception of particle, wave or ray properties. Here also, decoherence from external interactions can play a role but is not essential in deciding the extent of perception.

In the case of the detection of the effects of particle entanglement it is necessary to treat the ensemble entity as that corresponding to the many-particle wavefunction and its quantum numbers. Incomplete extraction of the information encoded in the many-particle ensemble wavefunction, for example detection of only some of the particles or incomplete specification of vector variables, corresponds to an effective decoherence.

\section*{Acknowledgements}
The author is extremely grateful to Prof. J.M. Feagin for several years of close cooperation on the derivation and meaning of the imaging theorem.

 \appendix
 
 \section{The density matrix}
As in the discussion of decoherence by Zurek \cite{decohere1}  the time development of a one-dimensional single-particle  ensemble wavepacket is considered. The wavepacket is composed of two gaussians centred at $x = X_1$ and  $x = X_2$ with width such that there is essentially no overlap at $t=0$.
The initial state is then
\begin{equation}
\Psi(x,t=0) = (\pi\sigma^2)^{-1/4}\sum_{i=1,2} e^{-x_i^2/(2\sigma^2)} 
\end{equation}
where $x_i \equiv x - X_i$.
For $t > t_0$ this initial wavefunction propagates freely in time and has the exact form
\begin{equation}
\begin{split}
\Psi(x,t) = (\sigma^2/\pi)^{1/4}& \left(\sigma^2 + \frac{i\hbar t}{\mu}\right)^{-1/2}~\\ & \times \sum_{i=1,2} \exp{\left[-\frac{x_i^2}{2\left(\sigma^2+\frac{i\hbar t}{\mu}\right)}\right]} 
\end{split}
\end{equation}
The IT condition emerges in the limit of large times and distances. Large times corresponds to $\hbar t/\mu >> \sigma^2$. Then the spatial wavefunction assumes the IT form,
\begin{equation}
\begin{split}
\Psi(x,t) \approx & \left(\frac{\sigma^2}{\pi}\right)^{1/4}\left(\frac{\mu}{i \hbar t}\right)^{1/2}\\ & \times \sum_{i=1,2}e^{-(\mu x_i\sigma/(\sqrt 2\hbar t))^2} 
e^{i \mu x_i^2/(2\hbar t)} 
 \end{split}
\end{equation}
The IT limit giving the classical trajectory is such that $x_i$ and $t$ both become large but the ratio is a constant classical velocity. To emphasise this we introduce the momenta $p_i = \mu x_i/t$. We also define, as the width of the Gaussian in momentum space, $\tilde\sigma \equiv \hbar/\sigma$. Then we can simplify the asymptotic spatial wavefunction using 
\begin{equation}
(\mu x_i\sigma/(\sqrt 2\hbar t))^2 \equiv   p_i^2/(2\tilde\sigma^2)
\end{equation}
and the energy phases
\begin{equation}
\mu x_i^2/(2\hbar t) =  p_i^2t/(2\mu\hbar).
\end{equation}
The asymptotic spatial wavefunction is then,
\begin{equation}
\Psi(x,t) \approx\left( \frac{\mu}{i\sqrt\pi \tilde\sigma t}\right)^{1/2} \sum_{i=1,2} e^{-p_i^2/(2\tilde\sigma^2) + ip_i^2t/(2\mu\hbar)} 
\end{equation}
which looks exactly like a pair of free momentum gaussians propagating in time and corresponds to the 1D form of the IT of \eref{IT_coor}, with $dp_i/dx_i = \mu/t$ for free motion.

The diagonal element of the density matrix is defined as $\rho(x,x,t) = \Psi^*(x,t)\Psi(x,t)$ and is
\begin{equation}
\label{rhodiag}
\begin{split}
\rho(x,x,t) = & \frac{\mu}{\sqrt\pi \tilde\sigma t} \sum_{i=1,2} e^{-p_i^2/\tilde\sigma^2} \\ &+ 2 \cos{[(p_1^2 - p_2^2) t/(2\mu\hbar)]} ~e^{-(p_1^2 + p_2^2)/(2\tilde\sigma^2)}
\end{split}
\end{equation}

The off-diagonal density matrix is defined as $\rho(x, x',t) = \Psi^*(x,t)\Psi(x',t)$ and consists of four terms,
\begin{equation}
\label{rhooffdiag}
\begin{split}
\rho(x, x',t) = & \frac{\mu}{\sqrt\pi \tilde\sigma t} \sum_{i,j=1,2} e^{-(p_i^2 +  p_j^{\prime 2})/(2\tilde\sigma^2)}e^{-i(p_i^2 -  p_j^{\prime 2})t/(2\mu\hbar)}
 \end{split}
\end{equation}
At $t=0$ this gives rise to four gaussian peaks, as in Ref.\cite{decohere1}. It reduces to the diagonal element when $p_i = p'_i$, i.e. $x = x'$ as it should. 

One sees that the diagonal matrix element shows two peaks at $p_i = 0, p_2 = 0$ or  equivalently  $x=X_1,x=X_2$. There is also an interference term. In the off-diagonal element there are four peaks, with the two additional peaks at $x'=X_1$ and $x'=X_2$. These also contain oscillatory phase factors giving interference.

Clearly, to observe interference effects the temporal resolution must typically be less than one oscillation, i.e. $t < 4\mu\pi/(p_1^2 - p_2^2)$. If we take the two peaks to be separated by 1a.u., then in atomic units we have $ t< 4\pi \approx 10^{-16} secs.$. However, typical resolutions are nanosecs., that is seven orders of magnitude larger than this. If the resolution is $\delta t \equiv \tau$ then the measurement must be integrated over this time period. Typically the oscillatory terms will then give, omitting constants
\begin{equation}
\int_{-\tau/2}^{\tau/2} e^{i(p_1^2 - p_2^2) t}~ dt \approx \delta(p_1^2 - p_2^2).
\end{equation}
and similarly for the off-diagonal element when $p_i$ is replaced by $p_i'$.
In other words, the oscillations will average to zero under low resolution of measurement on an atomic time scale.  From Eq.\ref{rhodiag} this implies that the density matrix will  exhibit only  two diagonal gaussian peaks for such measurements, 
\begin{equation}
\rho(x,x,t) =  \frac{\mu}{\sqrt\pi \tilde\sigma t} ~(e^{-p_1^2/\tilde\sigma^2} + e^{-p_1^2/\tilde\sigma^2})
\end{equation}
with $p_i = \mu(x-X_i)/t$.
For the off-diagonal elements, from \eref{rhooffdiag}, all the terms will average to zero under normal time resolution to give zero off-diagonal elements..
 This is exactly the limit, elimination of off-diagonal density matrix elements, given by Zurek \cite{decohere1} as the classical limit and resulting from time propagation in the presence of an interacting environment.  However, we emphasise again that the wavepackets on the diagonal are spreading and only in the limit that particles are macroscopically massive can this be ignored to give localised single particles as envisaged in \cite{Schloss1}. The SP picture is assumed. 
 
 By contrast the IT proves that classicality emerges from unitary
 Hamiltonian propagation under low temporal resolution, in that the density matrix then has only two diagonal peaks . Quantum coherence is lost except where the temporal and spatial resolution are extremely high.
 The peaks represent an ensemble of classical particles moving on classical trajectories, centred around the two centre-of-mass trajectories, and distributed according to the initial Gaussian momentum wavefunction.

\end{document}